\documentclass[sigconf,nonacm]{acmart}
\usepackage{caption}
\usepackage{subcaption}
\usepackage{multirow}
\usepackage{xcolor}
\usepackage{graphicx}
\usepackage{epstopdf}
\usepackage{graphics}
\usepackage{url}
\usepackage{comment}
\usepackage{duckuments}
\AtBeginDocument{%
  \providecommand\BibTeX{{%
  \normalfont 
  \kern-0.5em{\scshape i\kern-0.25em b}\kern-0.8em\TeX}
  }}

\begin{document}

\title{Quantifying Outlierness of Funds from their Categories using Supervised Similarity}

\author{Dhruv Desai}
\email{dhruv.desai1@blackrock.com}
\affiliation{%
\institution{BlackRock, Inc.}
\city{New York}
\state{NY}
\country{USA}
}
\author{Ashmita Dhiman}
\email{ashmita.dhiman@blackrock.com}
\affiliation{%
\institution{BlackRock, Inc.}
\city{Gurgaon}
\state{HR}
\country{India}
}
\author{Tushar Sharma}
\email{tushar.sharma1@blackrock.com}
\affiliation{%
\institution{BlackRock, Inc.}
\city{Gurgaon}
\state{HR}
\country{India}
}
\author{Deepika Sharma}
\email{deepika.sharma@blackrock.com}
\affiliation{%
\institution{BlackRock, Inc.}
\city{New York}
\state{NY}
\country{USA}
}
\author{Dhagash Mehta}
\email{dhagash.mehta@blackrock.com}
\affiliation{%
\institution{BlackRock, Inc.}
\city{New York, NY}
\country{USA}}

\author{Stefano Pasquali}
\email{stefano.pasquali@blackrock.com}
\affiliation{%
\institution{BlackRock, Inc.}
\city{New York, NY}
\country{USA}}

\renewcommand{\shortauthors}{Desai et al.}

\begin{abstract}
Mutual fund categorization has become a standard tool for the investment management industry and is extensively used by allocators for portfolio construction and manager selection, as well as by fund managers for peer analysis and competitive positioning. As a result, a (unintended) miscategorization or lack of precision can significantly impact allocation decisions and investment fund managers. Here, we aim to quantify the effect of miscategorization of funds utilizing a machine learning based approach. We formulate the problem of miscategorization of funds as a distance-based outlier detection problem, where the outliers are the data-points that are far from the rest of the data-points in the given feature space. We implement and employ a Random Forest (RF) based method of distance metric learning, and compute the so-called class-wise outlier measures for each data-point to identify outliers in the data. We test our implementation on various publicly available data sets, and then apply it to mutual fund data. We show that there is a strong relationship between the outlier measures of the funds and their future returns and discuss the implications of our findings.
\end{abstract}



\maketitle

\section{Introduction}

Mutual funds and Exchange Traded Funds (ETFs), which we will collectively refer to as funds in this work, have become widely used investment vehicles globally across institutional and wealth investors. As of Q1 2023, Mutual funds and ETFs account for \$63 trillion in assets globally \cite{ici-fund-flows}.

Funds are constructed using a wide range of investment strategies from tracking market indices to generating alpha above a defined market benchmark. Fund managers often have an incentive to differentiate their product offerings within the sub-asset class resulting in differences in holdings and composition. Due to the range of different options available to investors, there arose a need to categorize funds for performance comparison and analysis. 
Certain third-party data vendors such as Morningstar \citep{morningstarcategorization} and Lipper \citep{lippercategory} as well as Preqin \citep{preqin-doc}, HFRI \citep{hfri-doc}, etc. created peer groups categorization systems to help investors compare and benchmark funds based on portfolio objectives, investment strategy and composition. 

The categorization system may have a hierarchical structure starting from a broad asset-class level categorization such as equity, fixed income,
real estate, mixed asset, commodity, etc., to more granular categories. Each fund is assigned a unique category at each level of hierarchy based on quantitative and qualitative assessment by the vendors. Funds in each category are considered peers and expected to be similar in construction and expected performance, and therefore, extensively used for peer group benchmarking and allocation decisions by institutional and wealth investors.   

The fund categorization systems could be prone to biases and it has been well-documented. Refs.~\cite{marathe1999categorizing,orphanides1996compensation,brown1997mutual,chen2021don,dibartolomeo1997mutual,elton2003incentive} have extensively reported evidence of this misclassification problem. This could be due to a few factors:
\begin{itemize}
    \item \textit{Information asymmetry and high cost of information acquisition}: Vendors frequently rely on self-reported summary data from fund managers on a periodic basis. These summary metrics do not always provide a complete and accurate view into the fund’s strategy.
    \item \textit{Complexity of portfolio constituent data}: Portfolio holding data, especially for fixed income funds and private markets, can make it difficult to independently monitor and verify fund information.
    \item \textit{Ambiguity, heterogeneity, and lack of sufficient granularity in categorization systems}: These three, along with any other irregularities in the classification system can result in a broad range of funds getting bucketed into the same category.
\end{itemize}

Often, misclassification can be beneficial to certain fund managers. For example if the fund appears to be less risky or appears to have better performance track record etc. to the investors, the fund manager can charge higher fee versus peers. Ref.~\cite{cooper2005changing} shared evidence that a change in the name of a fund as well as its stated style change may significantly increase net inflows to the fund.

Given the importance of fund categorization within asset management, misclassification can have broader consequences for investors that have been studied and quantified previously. Ref.~\cite{dibartolomeo1997mutual} argued that investment style misclassification may have a significant effect on investors’ ability to build diversified portfolios of funds, whereas Ref.~\cite{cooper2005changing} showed that change of name of a fund as well as stated style change may significantly increase net inflow to the fund. Ref.~\cite{bams2017investment} showed that misclassified funds significantly under perform well-classified funds in the long run. Ref.~\cite{chen2021don} argues that miscategorized funds have higher average risk as well as accompanying yields on their holdings than its category peers.

A few studies have employed machine learning (ML) techniques to investigate how similar or dissimilar funds are, and in turn to measure and correct for misclassification. In Ref.~\cite{marathe1999categorizing}, results from fund clustering were compared to vendor categories to capture any differences between the data-driven approach and the classification system. Refs.~\cite{haslem2001morningstar,sakakibara2015clustering,lamponi2015data,menardi2015double,vozlyublennaia2018mutual,kim2000mutual,castellanos2005spanish,moreno2006self,acharya2007classifying,lajbcygier2008soft,satone2021fund2vec} take a similar approach albeit with different clustering methodologies.

Across the studies, most clusters did not match the original categories completely. Some studies such as Ref.~\cite{haslem2001morningstar} claimed these inconsistencies may indicate flaws in categorization by vendors. In a rebuttal to Ref.~\cite{haslem2001morningstar}, Gambera, Rekenthaler and Xia argued that the drivers of the inconsistencies, including (1) the variables used for clustering were not exactly the variables used for their categorization by vendors, and (2) a categorization system is a classification system, so an unsupervised method such as clustering method should not be expected to reproduce it.

Indeed, the authors of Ref.~\cite{mehta2020machine} and\cite{vamvourellis2022learning} showed that using supervised ML techniques instead can help reproduce Morningstar and Lipper categorizations with high accuracy if the data used to train the model includes fund composition and self-reported investment objectives as input features and the categories as the target variable, and posing the problem as a supervised multi-class classification problem. This, in turn, shows that the existing systems are internally fairly consistent and rules-based.

Later, in Ref.~\cite{desai2021robustness}, the authors also reproduced categories as distinct clusters using a combination of distance metric learning and K-means clustering. This helped resolve the ongoing debate about reproducing categorization system using ML techniques while pointing out the inappropriate use of ML techniques in most of the previous works related to unsupervised clustering.

In the present work, we formulate the problem of quantifying `outliererness' first as a supervised distance metric learning using RF, and employ a distance-based outlier measure that yields the distance of a fund from the rest of the funds within its assigned category. Hence, the outlier measure provides a continuous measure to quantify the outlierness of a fund as opposed to the traditional approaches which decides miscategorization in binary fashion (i.e., only whether the fund is miscategorized with respect to its assigned category or not).
\vspace{-4mm}
\subsection{Our Contributions}
Our novel contributions to the existing literature in the present work are as follows:

\begin{enumerate}
	\item We pose fund miscategorization as a distance metric based outlier detection problem, and in turn device a continuous measure of outlierness of the funds with respect to their assigned category (as well as all other categories);
	\item We develop a Python library implementation for the proposed methodology, and benchmark our implementation on publicly available toy datasets;
	\item We then employ the outlier measure on the fund data and show that there is inverse correlation between the fund's outlier measure and its returns.
\end{enumerate}

We organize the remainder of the paper as follows: in Section \ref{sec:dis_metric}, we introduce the problem of fund categorization from the ML point of view and show that the problem is well-suited for a distance-based outlier detection methodology. In Section \ref{sec:data_description}, we provide details of the public datasets as well as the fund data used in our experiments. In Section \ref{sec:modeling_details}, we describe our methodology and experimental set up in details such that the experiments can be reproduced by other researchers provided the fund data is available to them. In Section \ref{sec:results}, we discuss our results, and finally we provide outlook and conclusions of the results in Section \ref{sec:conclusion}.

\section{Fund Miscategorization as a Distance Metric Learning Problem}\label{sec:dis_metric}

Identifying a fund that is misfit, or an outlier, with respect to its assigned category is clearly an outlier detection problem in the ML terminology. Outlier detection is a ubiquitous problem in many sub-areas of finance, and usually it is posed as an unsupervised learning problem or sometimes as a supervised learning problem if the ground truth labels for outlier data-points are available. In the present work, we implement and employ a supervised similarity (or, distance metric) learning based method that first learns the distance metric for the given data-set using appropriate labels, and then measures the outlierness of each data-point with respect to the rest of the data-points within the same class. That is, the outlier measure assigns a continuous value to each data-point with respect to its assigned true class, instead of merely a binary flag as an outlier or not, which in turn provides a way to systematically study the effects of fund miscategorization on the portfolio returns with respect to its peers.

\subsection{Supervised Similarity Learning}
Although similarity learning has found many applications in financial services, most of the research there has been focused on unsupervised methods such as $K$-means or other clustering methods where a specific distance metric such as Euclidean, Chebyshev, Minkowski, etc. is explicitly or implicitly supplied in the form of the chosen objective function. However, such a manually supplied distance metric may or may not be appropriate for the underlying data manifold. Moreover, here, the set of features to define similarity as well as the importance (i.e., weights) of each feature is also usually manually supplied.

Contrary to that, the similarity or distance metric learning (DML) methods algorithmically learn the distance metric (or, similarity metric) using the given data and labels in a supervised or semi-supervised fashion: instead of relying on manually supplied distance metrics, DML aims to algorithmically construct a distance metric from the given labeled data that can improve the performance of a distance based model for downstream tasks such as classification (e.g., K-nearest neighbor (K-NN) algorithm) or clustering (e.g., K-means) Ref.~\cite{10.5555/2968618.2968683,yang2006distance,kulis2012metric,DBLP:journals/corr/abs-1812-05944,pang2006pathway,resende2018survey,zhao2016propensity}.

To the best of our knowledge, a DML algorithm was used to learn similarity among mutual funds for the first time in Ref.~\cite{desai2021robustness}, though this DML algorithm ~\cite{10.5555/2968618.2968683} learns the global distance metric, i.e., not taking into account the local non-linearity and nuances of the data manifold. Moreover, this algorithm may not capture the distance metric over input space that is a mixture of categorical and numerical features. In Ref.~\cite{jeyapaulraj2022supervised}, following up from ~\cite{breiman-cutler-blog}, the authors argued that RF ~\cite{breiman2001random} can be viewed as a powerful method to learn similarity from complex datasets. To summarize, the advantages of using RF as a DML are: (1) RF is a non-parametric method and usually requires minimal hyperparameters tuning as well as data cleaning or preprocessing; (3) RF accepts both numerical and categorical as well as mixed types of input features; (3) it can scale well to relatively large datasets; (4) RF does not require reduction in dimensionality for the raw data; (5) RF learns a local distance metric (i.e., the distance metric that may vary as the location in the space if required) as opposed to a global distance metric.

Below we describe the process to compute similarity using RF.

\subsection{Similarity Learning using Random Forest}
First, we recall that Ref.~\cite{breiman-cutler-blog} proposed a specific way to extract similarity from a trained RF as follows: once $T$ trees are grown (i.e., RF is trained), we pass each of the data-points (from both training and out-of-bag (OOB) sets) through each tree of the forest. Then, if a pair of data-points falls in the same leaf node for a tree, we increase their similarity score (i.e., proximity) by one. Then, normalize the sum of the similarity scores between the pair over all the trees by $T$. Thus, the maximum possible proximity between a pair of data-points can be $1$ (i.e., the pair of data-points fall in the same leaf node in each of the $T$ trees) and the minimum can be $0$ (i.e., the pair never falls in the same leaf node in any of the $T$ trees in the forest). Mathematically, the proximity between data-point $i$ and $j$ is given by,
\begin{equation}
	Prox(i,j) = \frac{1}{T} \sum_{t=1}^{T}{I(j \in v_i(t))},
	\label{eq:prox}
\end{equation}
where $T$ is the number of trees in the forest, $v_i(t)$ contains indices of the data-points that end up in the same leaf node as $i$ in the tree $t$, and $I(.)$ is the indicator function. \footnote{We have also implemented and experimented with out-of-bag (OOB) proximities \cite{liaw2002classification} and Geometry-and-accuracy preserving (GAP) proximities \cite{lin2006random,rhodes2023geometry}. For the brevity of the presentation though, we provide results only using the proximity as defined by Eq.~(\ref{eq:prox}), and the results from other proximities will be discussed in a future work.}.

\subsection{Outlier Measure using Random Forest}
In general, outliers can be defined as the data points that are away from all other data points in the sample space. Thus, outliers will generally have small proximity to other data points. Ref.~\cite{breiman2001random} defined outlier measure for each data-point relative to the class it belongs to. Here, we generalize the outlier measure for a data-point relative to all the data-points of any class including its own class: first, we define the average proximity of each data point $i$ the data points in class $J$ as
\begin{equation}
	P^{J}(i) = \sum_{j\in cl(J), i\neq j} Prox^2(i,j).
\end{equation}
Here, $cl(J)$ denotes set of the data-points in class $J$, i.e., the sum in the formula runs over all the data-points within class $J$. If $J$ is the class of the data-point $i$, then the sum runs over all the data-points within the class except the data-point $i$.

Now, we can define the raw outlier measure for the $i$-th data point with respect to all the data points in its class $J$ as,
\begin{equation}
	P_{\mbox{raw}}^{J}(i) = \frac{n_J}{P^{J}(i)},
\end{equation}
where $n_J$ is the number of data-points in class $J$.
To arrive to the final outlier measure for the $i$-th data-point, we first find the median of the raw outlier measure, $\mbox{med}_{J}(P_{\mbox{raw}}^{J}(i))$, and their absolute deviation from the median, $\mbox{dev}_{J}(P_{\mbox{raw}}^{J}(i))$, both over all the data-points in class $J$ \textit{as well as the data-point $i$ if the ground-truth class of $i$ is different than $J$}. Then, the final outlier measure is defined as:
\begin{equation}
	O^{J}(i) = \frac{P_{\mbox{raw}}^{J}(i) - \mbox{med}_{J}(P_{\mbox{raw}}^{J}(i))}{\mbox{dev}_{J}(P_{\mbox{raw}}^{J}(i))}.
\end{equation}
The larger the value of $O^{J}(i)$ for the $i$-th data-point, the farther it is from the data-points of class $J$.

\subsection{Visualization}
We use two different ways to visualize the results: class-wise scatter plot of outlier measures; and Multi-Dimensional Scaling (MDS).

For the class-wise scatter plot of outlier measures, we plot the outlier measures for the data-points in each target class where the classes are denoted on the y-axis to show how far or close individual data-points are from the rest of the data-points in the corresponding class on the x-axis.

To have a two-dimensional visual representation of the distance matrix, we utilize the technique of MDS \cite{hout2013multidimensional}. With the input of a symmetric proximity matrix, output of MDS is a map that exhibits the relationships among data-points where similar data-point are located closer to each other and dissimilar data-points are located proportionately placed further apart.

For the proximities among all pairs of data-points in the dataset generated from RF, the distance matrix is defined as $1-$ Proximity, where the Proximity is an $n\times n$ matrix whose entries are one of the pair-wise proximity measure as defined above and $n$ is the total number of data-points in the dataset.

\section{Data Description}\label{sec:data_description}
In this Section, we describe details of public datasets as well as fund data and the data preprocessing procedure employed in this work.

\subsection{Toy Datasets}
To benchmark our implementation, we test it on various toy datasets for classification task from the University of California Irvine (UCI) repository\cite{blake1998uci}: Iris flower data, MNIST digits, Wine, Breast cancer and car-evaluation. A brief description of toy classification datasets used in this paper are as in shown Table \ref{tab:data-summary}.

\begin{table}
    \small
	\begin{tabular}{l l l l}
		\toprule
		Data            & No. of Classes & Features & No. Obs. \\ 
    \midrule
    Iris            & 3           & 4 (Numerical)       & 150     \\
		Digits          & 10          & 64 (Numerical)      & 1797    \\
		Wine            & 3           & 13 (Numerical)      & 178     \\
		Breast cancer   & 2           & 30  (Numerical)     & 569     \\
		Car Evaluation  & 4           & 6 (Categorical)           & 1728    \\
    Funds           & 119         & 373 (Mixed)         & 10429   \\
		\bottomrule
	\end{tabular}
	\caption{Summary for Classification datasets}
  \label{tab:data-summary}
\vspace{-6mm}
\end{table}

\subsection{Fund Data}
We sourced our data for funds from Morningstar Data warehouse data feed. Since Morningstar Categories (the target variable) are based on funds' portfolio composition, we chose features from the feed which would describe the same. This dataset provides various levels of aggregation breakdowns which would help explain the funds' portfolio composition. Individual feature aggregation often breakdown into different subgroups and within each subgroup this dataset provides the percentage breakdown of fund invested into this subgroup. For example, when looking at a funds equity sector breakdown, there are a total of eleven sectors and the dataset gives the percentage breakdown of funds investment into each sector. These aggregations vary across different attributes listed in Table \ref{tab:fund_data}. Arguably, these aggregations can be considered as better features to understand a fund's portfolio composition as compared to the fund's security-level holding information as the former provide generalization across different funds and asset classes.

\begin{table}[ht]
    \centering
    \small
    \begin{tabular}{l l l}
    \toprule
         Feature Group          & Total Subgroups  & Feature type\\ 
         \midrule
         Benchmark Features     & 3   & Categorical \\
         Asset Allocation       & 14  & Numerical \\
         Asset Class            & 7   & Categorical \\ 
         Bond Region            & 16  & Numerical \\
         Bond Primary Sector    & 34  & Numerical  \\
         Bond Secondary Sector  & 12  & Numerical \\
         Muni Sector            & 55  & Numerical  \\
         Capital Breakdown      & 5   & Numerical  \\
         Coupon Range Breakdown & 42  & Numerical \\
         Stock Sector           & 11  & Numerical \\
         Maturity Range         & 13  & Numerical \\
         Region Breakdown       & 16  & Numerical  \\
         Stock Type             & 9   & Numerical  \\
         Style Box              & 9   & Numerical  \\
         Credit Quality         & 8   & Numerical  \\
         Country Breakdown      & 119 & Numerical \\
         \bottomrule
    \end{tabular}
    \caption{Fund data attributes}\label{tab:fund_data}
    \vspace{-8mm}
\end{table}

For the scope of this work we used data for December 2018 (which is available beginning of January 2019) to compute the outlier scores and the returns of all the funds for the three-year time-period starting from January 2019 to December 2021 to compute regression between outlier scores and returns. We have chosen these time-periods for the availability of the data. Our universe consisted of all U.S. domiciled oldest-share-class, open-end mutual funds and ETFs(\~10.5K funds). There were a total of hundred and nineteen Morningstar Categories covered in this dataset. The class distribution was imbalanced, where the largest class had ~500 samples and the smallest class had only 10 samples. To account for this imbalance, we used stratified sampling and took advantage of RF algorithm which automatically up scales smaller classes. In case of missing values, since these features describe the percentage of funds portfolio invested in the given aggregation, we imputed them with zero, i.e., the fund has no investments in that aggregation.

\section{Modeling Details}\label{sec:modeling_details}

Below, we describe details on our modeling methodology.

\subsection{Implementation Details}
We used scikit-learn \cite{scikit-learn} for all the computation in this work including except for the RF proximity computations which we implemented ourselves in Python.

There are a few implementations of the RF proximity measures in the \textsf{R} language, such as one from Breiman and Cutler themselves \cite{RF-package-breiman}. Another recent library \cite{Moon-RFGAP-package} for proximity computations in \textsf{R} is from the authors of Ref.~\cite{rhodes2023geometry}.

However, to the best of our knowledge, there is no rigorous Python library to compute various RF proximities. Current RF implementation in Python package scikit-learn \cite{scikit-learn} does not provide explicit functions to calculate proximity or outlier measure either, though this package does expose all its attributes which were used to build the individual trees in the ensemble of the trained RF. We developed our Python implementation of calculating RF proximites on top of the scikit-learn package. Furthermore, we extended our implementation to calculate out-of-bag proximity and Geometry-and-accuracy (GAP) proximity, though we will compare and contrast the different proximities in a future work. In addition, we implemented methods to calculate outlier measure based of these proximities. We plan to make this package public in the near future. 

We have also implemented MDS using Scikit-learn's MDS Class by passing the pre-computed distance matrix.

\subsection{Training-testing split (stratified)} To prevent the algorithm from over-fitting, a training-testing split of $80-20$, respectively, with stratification to ensure that train and test split approximately have the same distribution of each target class as the complete set.

\subsection{Balanced Class weight} 
Since the dataset is highly imbalanced, we used balanced class weights to train the algorithm, i.e., the class-weights are then inversely proportional to their frequencies.

\begin{table}[htbp]
    \centering
    \begin{tabular}{l  l}
         \toprule
         Hyperparameter & Range  \\ 
         \midrule
        Number of trees & 100-1000 (step size 100) \\
        Max depth & 5-50 (step size 5) and till pure leaf node \\
        Max features & Sqrt, log2 of total features \\
        Split criterion & Gini, Entropy and Log loss \\
        \bottomrule
    \end{tabular}
    \caption{Hyper-parameter ranges for grid search}\label{tab:hyper_param_range}
    \vspace{-8mm}
\end{table}

\subsection{Hyperparameter Optimization and Cross-validation} 
To tune hyperparameters of RF, we used stratified $5$-fold cross-validation on the training dataset, where the training set is spilt into $5$ smaller sets with all the folds having the same percentage of samples of each target class as that in the training data. We performed grid search for number of trees, max depth, maximum number of features considered for best split and split criterion. Table \ref{tab:hyper_param_range} provides additional details on ranges of different parameters used for grid search.

\subsection{Evaluation Metrics}
We have used various evaluation metrics for both the stages, i.e., the classification task and the similarity computation using RF proximities, as follows.

\subsubsection{Evaluation Metrics for the Classification Task}
Since our data is highly imbalanced, we used not only accuracy but also $F1$ score, the weighted micro $F1$ score, the weighted macro $F1$ score and micro and macro area under the receiver-operating characteristic curve (AUC-ROC). These evaluation metrics also provide the extent of miscategorization in absolute sense, i.e., whether the true category of a fund can be reproduced by the supervised classification algorithm.

\subsubsection{Evaluation Metrics for the Utility of the Outlier Measure}
To evaluate the utility of the outlier measure, we looked for evidence of relationship between the outlier score of a fund and its future returns: if a fund screens as an outlier compared to its peers in a Morningstar category, then we expect a weak relationship between the fund’s future returns and the respective Morningstar category benchmark over the same time-period.

Here, we first calculate the outlier score based on the fund data as of Dec 2018. Alongside, we sourced the returns of all the funds for the three-year time-period starting from January 2019 to December 2021. Additionally, we sourced the aggregate returns of the respective Morningstar category benchmark as well. We then split these funds into quartiles of outlier measures (based on December 2018 cross-section) within each of their Morningstar categories. The funds with the highest outlier score fall into the fourth quartile while funds with the lowest outlier score fall into the first quartile. We then ran a Linear Regression model between the Morningstar category benchmark returns as an independent variable and the fund returns as the dependent variable. This was done to record the $R^2$ of the regression to measure the magnitude of the variation of fund’s returns that was explained by the returns of the benchmark. In theory, a fund with $R^2$ value closer to $100\%$ would indicate that most of the movements in the fund can be explained by movements in the Morningstar category benchmarks. Later we looked at the distribution of the R-Squared values, including the $25^{th}$ percentile, median and $75^{th}$ percentile, for funds in each of the four quartiles of the outlier score.

\subsection{Computational Efforts}
We performed our computations using standard 32GB RAM machine with 3 cores. The computation of training the RF for funds data took ~5 minutes, that of extracting the similarity matrix out of the train RF took ~20 minutes, and computing outlier measure for all the data-points took ~4 minutes.

\section{Results}\label{sec:results}
In this Section, we describe the results from our experiments.
\subsection{Results from the Supervised Classification Phase}
As a first step, we trained RF for the toy as well as funds data on the training split using 5-fold cross-validation, and tested the model on the test dataset. The results for the best hyperparameter values for the respective datasets are recorded in Table \ref{table:results-classification} for the test data. 

The performance of the RF as measured by accuracy as well as macro and micro $F1$ scores and AUC-ROC for all the datasets are quite high. This is not surprising even for the funds data as fund categorization is a rule based system and internally fairly consistent as noted in Ref.~\cite{mehta2020machine}. The remaining misclassified funds are essentially the funds that the trained RF predicted a different category than the one assigned by Morningstar and hence these are clearly potential outliers by the usual interpretation of classification results.

\begin{table}[ht]
	\centering
    \small
	\begin{tabular}{l l l l l l}
    \toprule
    \textbf{Data} & \textbf{Acc} & \multicolumn{2}{l}{\textbf{F1-Score}} & \multicolumn{2}{l}{\textbf{AUC-ROC}} \\ \cline{3-6}
           & \textbf{}    & \textbf{Micro} & \textbf{Macro} & \multicolumn{1}{l}{\textbf{Micro}} & \textbf{Macro} \\	
           \midrule
        \textbf{Iris}           & 0.97 & 0.97 & 0.96 & 0.99 & 1.00 \\
		\textbf{Digits}         & 0.98 & 0.98 & 0.98 & 0.99 & 0.99 \\
		\textbf{Wine}           & 1.00 & 1.00 & 1.00 & 1.00 & 1.00 \\
		\textbf{Breast Cancer}  & 0.94 & 0.94 & 0.93 & 0.99 & 0.99 \\
		\textbf{Car Evaluation} & 0.96 & 0.96 & 0.93 & 0.99 & 0.99 \\
		\textbf{Funds}          & 0.96 & 0.96 & 0.96 & 0.99 & 0.99 \\
		\bottomrule
	\end{tabular}
	\caption{Classification results for the test data.}
	\label{table:results-classification}
  \vspace{-8mm}
\end{table}

\subsection{Outlier Measures}
After showing satisfactory performance of the RF on the classification task for the given datasets, here we present results for the outlier measure.

\begin{figure*}[htbp!]
\centering 

\begin{subfigure}{0.33\textwidth}
  \includegraphics[width=\linewidth, height=1.0\textwidth]{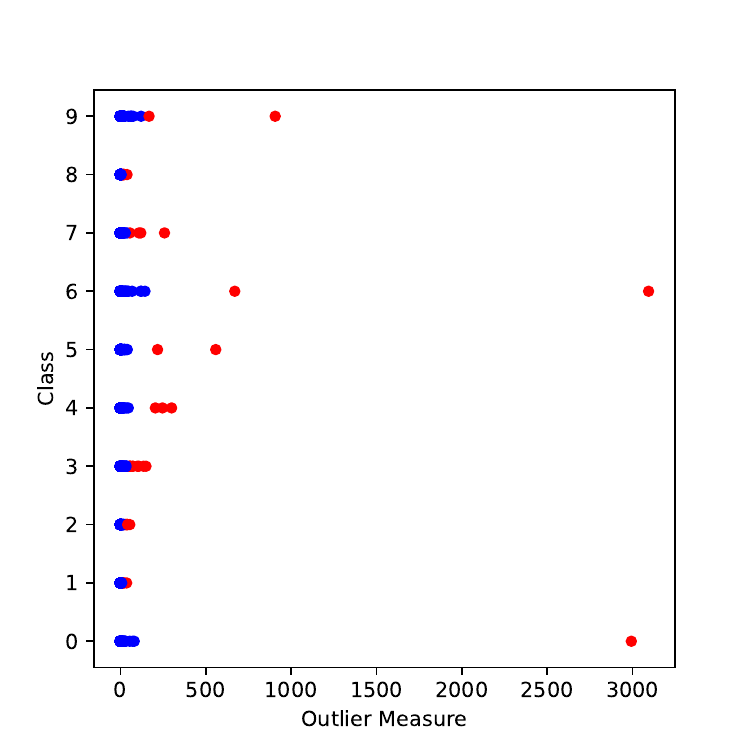}
  \caption{Digits}
  \label{fig:digit_outlier}
\end{subfigure}\hfil 
\begin{subfigure}{0.33\textwidth}
  \includegraphics[width=\linewidth, height=0.9\textwidth]{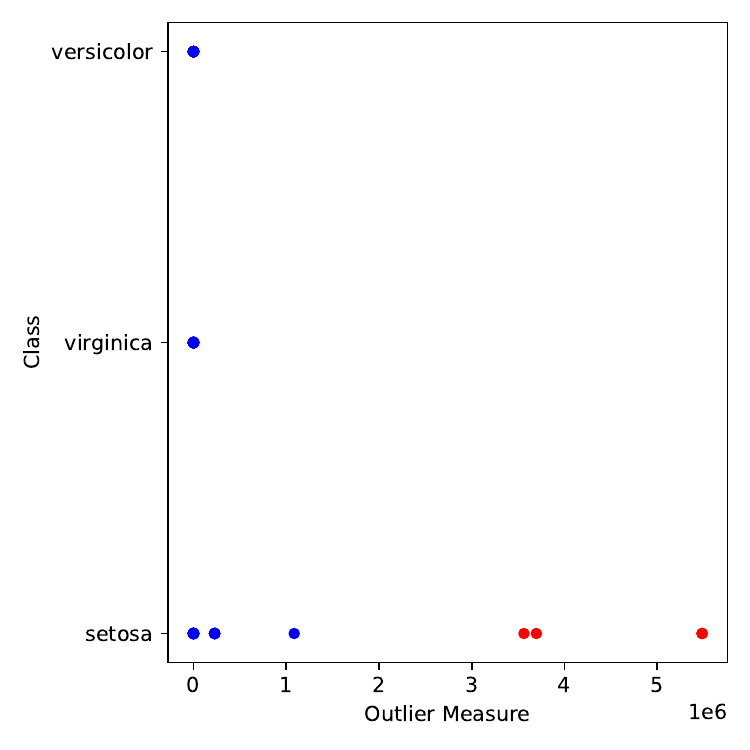}
  \caption{Iris}
  \label{fig:iris_outlier}
\end{subfigure}\hfil 
\begin{subfigure}{0.33\textwidth}
  \includegraphics[width=\linewidth, height=0.9\textwidth]{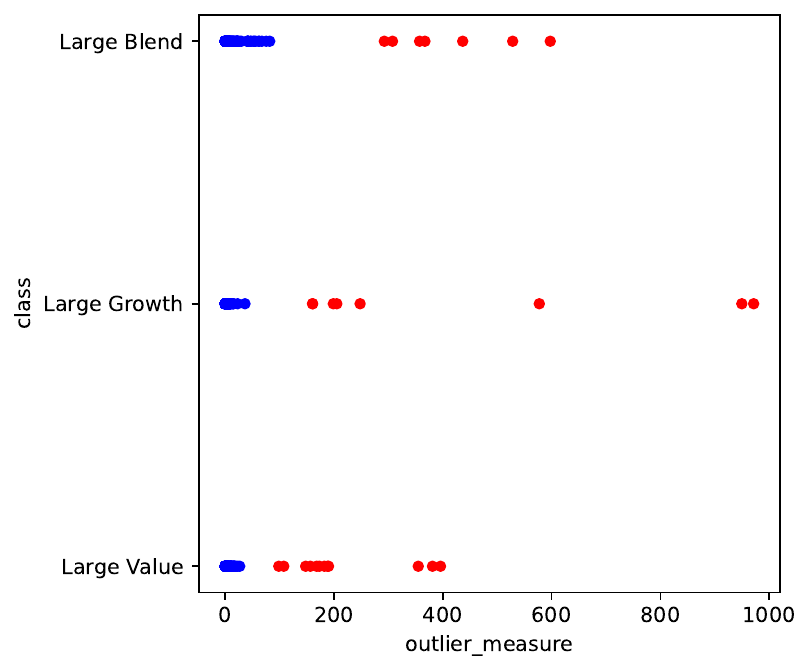}
  \caption{Funds \footref{mstar_cright}}
  \label{fig:fund_outlier}
\end{subfigure}
\caption{Outlier measure scatter plot with respect to their assigned classes.}
\label{fig:outlier_measure}
\end{figure*}

\subsubsection{Outlier Measure for the Within Class}
Figures (\ref{fig:digit_outlier})-(\ref{fig:fund_outlier}) show the distribution of outlier measure with respect to the rest of the data-points in its assigned class. Note that in Figure (\ref{fig:fund_outlier}), we restrict the visualization only to three Morningstar categories (US Large Blend, US Large Growth and US Large Value) for the brevity of presentation. For the same reason we will also focus our analysis to these three categories for the rest of the paper. In all these figures, we use the threshold of $2\times \mbox{(standard deviation)}$ of the outlier measure within the respective class beyond which the corresponding data-point is flagged as an outlier for the respective class. The outliers are colored red to distinguish them from the normal data-points.

\begin{figure}[htbp]
    \centering
     \includegraphics[width=0.75\linewidth]{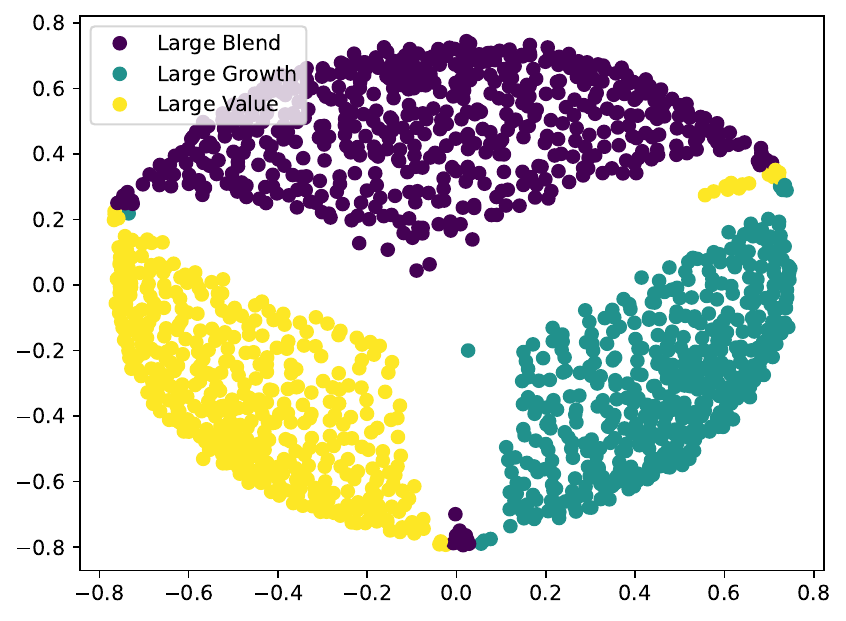}
    \caption{MDS plot for funds from three categories.}\label{fig:MDS}
    \vspace{-4mm}
\end{figure}

\subsubsection{Outlier Measure for Other Categories than the Assigned One}
\begin{figure*}[ht]
  \includegraphics[width=\linewidth, height=0.5\textwidth ]{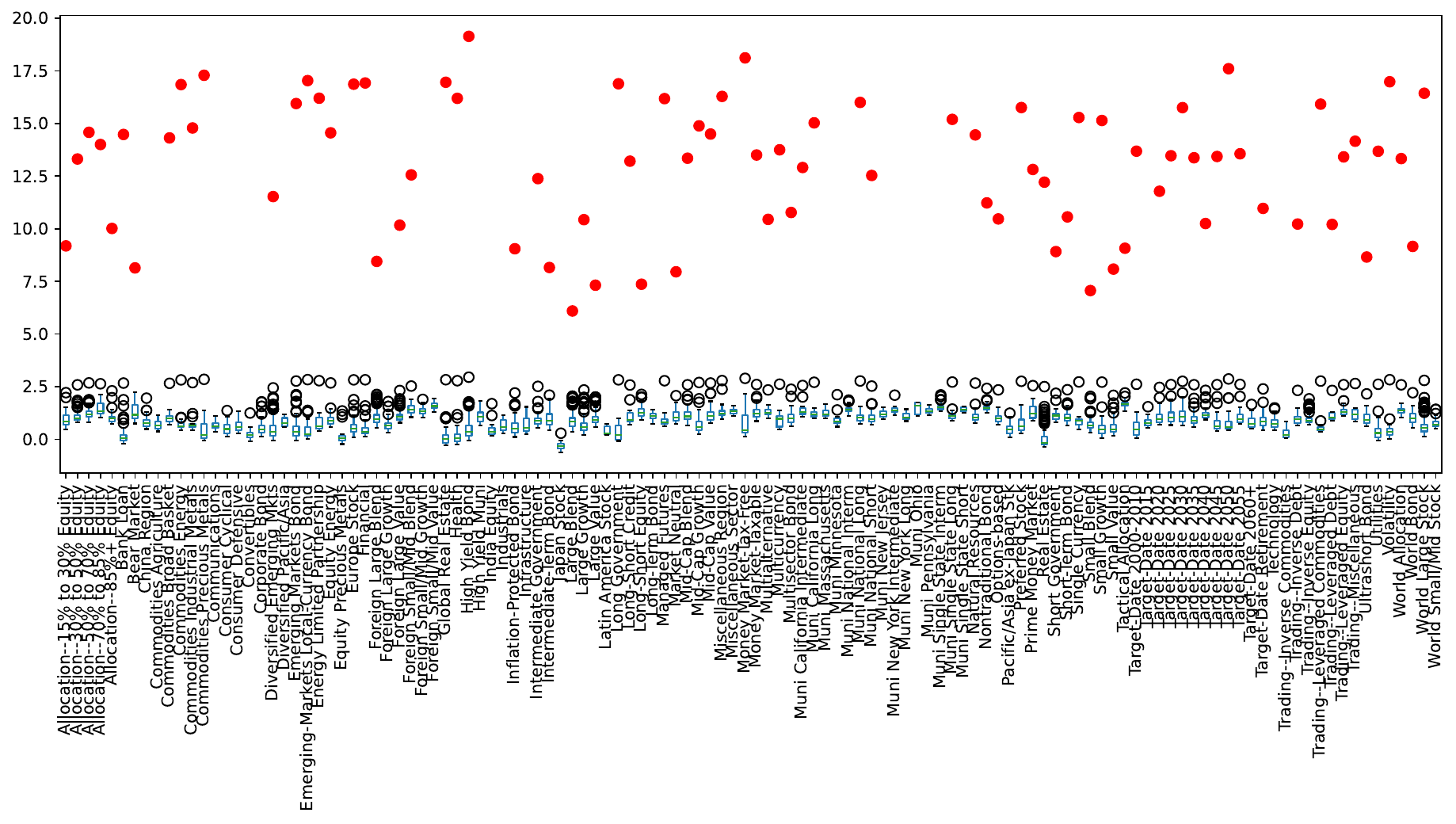}
  \caption{Using outlier score to detect novelties \footref{mstar_cright}}\label{fig:novelty}
\end{figure*}

We can also compute outlier measure for a data-point with respect to any of the other class than its ground-truth class. To that extent, for the funds data, for those data-points that are tagged as outliers with respect to their Morningstar categories, we now compute outlier measure with respect to other categories than their assigned category. Using this approach we can identify \textit{novelties} in the dataset, i.e., a data point which is an outlier to all other classes. We show one such example in Figure (\ref{fig:novelty}) which captures the log of outlier measure for the given point against all other classes in the dataset.

In addition, Figure (\ref{fig:MDS}) also shows the MDS plot for the three categories. Although it is difficult to extract quantitatively useful information from a reduced representation of the high-dimensional data, the figure at least visually exhibits that the funds deemed outliers fall neither in the clusters of data-points of their assigned categories.

\subsection{Miscategorized Funds and Outlier Measures}
One would wonder if there is any relationship between the miscategorization of data-points by RF for the classification task and the outliers flagged by the outlier measure. Table \ref{tab:misclas_out_ovrlap} shows the overlap between the miscategorized data-points and outliers.

It may appear that the outlier measure and RF miscategorization yield the same information, though, first, the RF miscategorization results are related to the probability thresholds on the decision function for (binary or multi-class) classification. The amount of data-points to be miscategorized by RF depends on the this probability threshold which can also be varied to obtain different results, the same way as the threshold on the outlier measures. An objective method to select an optimal threshold for unsupervised anomaly detection methods is still an active area of research with lack of conclusive results \cite{ma2023need}.
\smallskip

\begin{table}[htbp!]
  \centering
  \small
  \begin{tabular}{l l l}
    \toprule
    \textbf{Data}           & \textbf{No. Misclassification} & \textbf{No. outliers} \\
    \midrule
        \textbf{Iris}           & 1                              & 9   \\
		\textbf{Digits}         & 7                              & 48  \\
		\textbf{Wine}           & 0                              & 9    \\
		\textbf{Breast Cancer}  & 7                              & 11   \\
        \textbf{Car}            & 13                             & 67   \\
		\textbf{Funds}          & 104                            & 423  \\
		\bottomrule
  \end{tabular}
  \caption{Overlap between outliers and misclassification.}
	\label{tab:misclas_out_ovrlap}
  \vspace{-8mm}
\end{table}

\subsection{Fund Returns and Outlier Measures}
\begin{figure*}[htbp!]
\centering 
\begin{subfigure}{0.45\textwidth}
  \includegraphics[width=\linewidth]{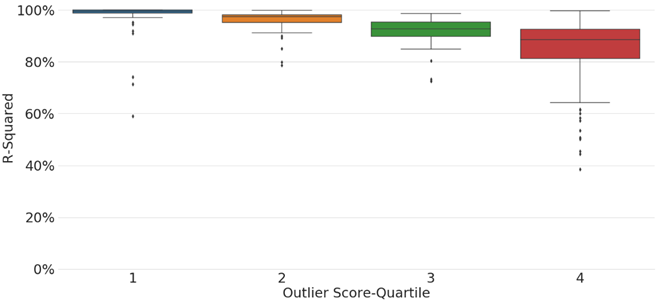}
  \caption{Category: Large Blend}\label{fig:lrg_bld_r2}
\end{subfigure}\hfil 
\begin{subfigure}{0.45\textwidth}
  \includegraphics[width=\linewidth]{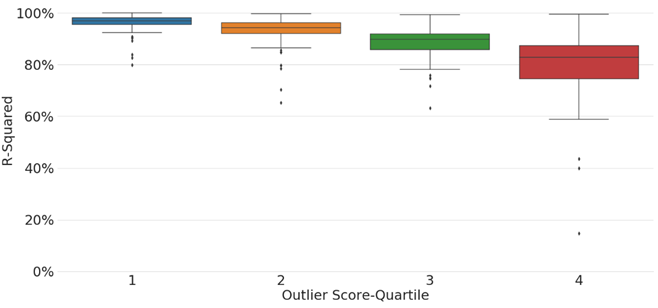}
  \caption{Category: Large Growth category}\label{fig:lrg_grw_r2}
\end{subfigure} 

\begin{subfigure}{0.45\textwidth}\hfill
  \includegraphics[width=\linewidth]{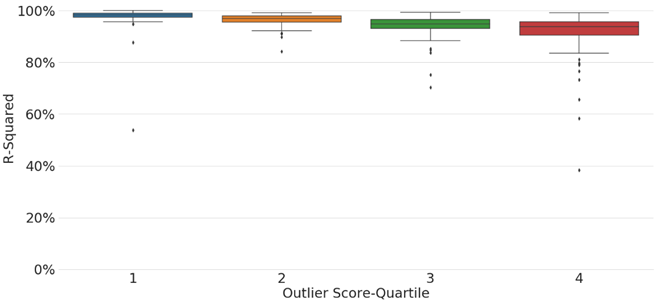}
  \caption{Category: Large Value}\label{fig:lrg_val_r2}
\end{subfigure}\hfil
\caption{Explained $R^2$ over quartiled outlier score for chosen categories \footref{mstar_cright}}\label{fig:explained_r2}
\vspace{-6mm}
\end{figure*}

We conducted the evaluation exercise across the three Morningstar categories – US Large Blend, US Large Growth and US Large Value. The respective benchmarks used for the three categories are S$\&$P 500, Russell 1000 Growth and Russell 1000 Value. The number of funds included for each of the three categories are 454, 373, and 340 funds respectively. We summarize our results in Figures (\ref{fig:explained_r2}).

Across the three categories, we observed a strong negative relationship between the outlier score quartile and the $R^2$ of each group. For the US Large Cap Blend category, we noticed that the median $R^2$ of funds in the first quartile of outlier score stood at $99\%$, implying that the S$\&$P 500 benchmark explained most of the movements in the underlying funds. However, the median $R^2$ for funds in the fourth quartile of the same category fell to $89\%$ indicating a weaker relationship with the S$\&$P 500 compared to funds in the first quartile. We noted numerous outliers $(Q1 - 1.5\, (\mbox{Inter-quartile range}))$ in the fourth quartile category as well, with roughly $10\%$ of the funds falling outside the box plot with $R^2$ as low as $40\%$. We observed similar patterns across other Morningstar categories – US Large Growth and US Large Value with key statistics summarized by box plots in Fig. \ref{fig:explained_r2}.
In summary, as the fund is identified as an outlier by the model, less of its future returns are explained by the respective Morningstar category benchmark.
\footnote{\label{mstar_cright} \copyright 2023 Morningstar. All Rights Reserved. The information contained herein: (1) is
proprietary to Morningstar and/or its content providers; (2) may not be copied or distributed; and
(3) is not warranted to be accurate, complete or timely. Neither Morningstar nor its content
providers are responsible for any damages or losses arising from any use of this information. Past
performance is no guarantee of future results.
}

\vspace{-2mm}
\section{Conclusion}\label{sec:conclusion}
Due to its influence on the financial decisions by the investors, miscategorization of funds by mutual fund categorizations may have severe implications for the end investors as well as for the asset management industry in general. In this work, we formulated the problem of identifying miscategorized fund as a supervised anomaly detection problem, and used distance metric learned by the RF proximities while learning the mapping between portfolio composition data of funds and their assigned Morningstar categories. We then implemented and employed an outlier measure which provides a continuous measure to determine the distance between the given fund to the rest of the funds in a category (the fund's assigned Morningstar category or else). We back-tested our methodology on funds' relative performance to demonstrate that the outlier measure is indeed negatively correlated with the relative returns of funds with respect to their peers within the assigned category.

Our work has potential applications which are directly relevant for the investment management industry:

\textit{Capturing Style drift}:- Style drift occurs when a mutual fund's actual and the declared investment style differs. When Style drift manifests itself in a mutual fund, the investment information that is publicly available for the fund becomes misleading. The deviated fund tends to become another product of different risk-return profile that is not aligned with the investor’s initial investment goal. Our work can help measure Style drift on a regular basis using the latest available positioning data for the funds. Categories provided by third-party vendors do not change often; however, using our approach we can help determine if a fund has moved away far from its peers using the outlier score – hence capturing the style drift.

\textit{Peer group benchmarking}:- Peer group benchmarking is of critical relevance in the fund management industry. Other than investment returns, peer group ranking is one of the key parameters used by investors/consultants to hire or fire any manager from their portfolio. Our approach can help determine if the funds included in the peer group are indeed relevant peers to be compared against for a fund. Our work can help to weed out funds which should not be a part of the comparison peer group - hence leading to better investment decisions.

\section*{Acknowledgement}
The views expressed here are those of the authors alone and not of BlackRock, Inc.
\bibliographystyle{ACM-Reference-Format}
\bibliography{sample-base}
\end{document}